\documentclass[a4paper,twocolumn]{article} 
\usepackage{amsmath,amssymb,graphicx,cite}
\title{On a discrete optimal velocity model and its continuous and ultradiscrete relatives\footnote{This article is submitted to JSIAM Letters.}}
\author{
Daisuke Takahashi\footnote{daisuket@waseda.jp} and Junta Matsukidaira\\
}
\date{}
\begin{document}
\maketitle
\abstract{We propose a discrete traffic flow model with discrete time.  Continuum limit of this model is equivalent to the optimal velocity model.  It has also an ultradiscrete limit and a piecewise-linear type of traffic flow model is obtained.  Both models show phase transition from free flow to jam in a fundamental diagram.  Moreover, the ultradiscrete model includes the Fukui--Ishibashi model in a special case.}
\section{Introduction}
  There are various models of different levels of discreteness to analyze the traffic congestion\cite{chowdhury}.  Macroscopic model is defined by a partial differential equation based on fluid dynamics and it describes a traffic flow by the motion of continuous media.  For example, Musha and Higuchi used the Burgers equation to describe a fluctuation of traffic flow\cite{musha}.\par
  System of ordinary differential equation (ODE), coupled map lattice (CML) and cellular automaton (CA) are often used as microscopic model to describe each vehicle motion directly.  About ODE models, time $t$ and vehicle position $x$ are continuous, and vehicle number $k$ is discrete.  CML is similar to ODE but time is discretized\cite{tadaki}.  All dependent and independent variables are discrete for CA models.  For example, Nagel--Schreckenberg model\cite{nagel}, elementary CA of rule number 184 (ECA184)\cite{wolfram}, Fukui--Ishibashi (FI) model\cite{fukui} and slow-start model\cite{takayasu} are known as effective traffic model.  Though evolution rule of CA model is simple due to its discreteness, a mechanism of congestion formation is presented sharply.\par
\begin{figure}
\begin{center}
  \includegraphics[scale=0.7]{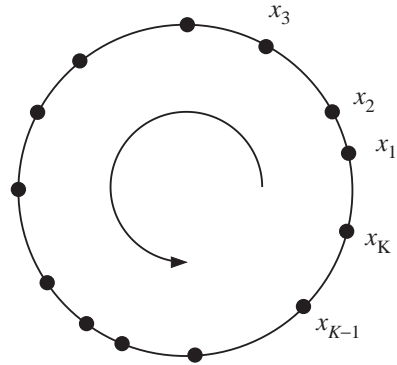}
\end{center}
  \caption{Circuit and vehicles.}
  \label{fig:circuit}
\end{figure}

  Bando et al proposed a noticeable ODE model\cite{bando}.  The model is now called `optimal velocity model' (OV model) and is defined as following.  Assume a finite number of vehicles moving on a one-way circuit of single lane as shown in Fig.~\ref{fig:circuit}.
  The length of the circuit is $L$ and total number of vehicles is $K$.  Introduce a one-dimensional coordinate along the circuit with an appropriate origin.   Define $x_k(t)$ by a position of vehicle with vehicle number $k$ ($k=1$, 2, $\cdots$, $K$) at time $t$.   The vehicle number is given sequentially to each vehicle as the preceding one has a larger number.  Note that the preceding vehicle of $k=K$ is $k=1$.  Then the evolution equation on $x_k(t)$ is
\begin{equation}  \label{ov}
  \ddot x_k=A\{V(x_{k+1}-x_k)-\dot x_k\},
\end{equation}
where $A$ is a constant representing a driver's sensitivity and $V(\Delta x)$ is an optimal velocity representing a desired velocity of a driver with a distance $\Delta x$ between his vehicle and the vehicle ahead.   The acceleration of $k$-th vehicle is determined by (\ref{ov}) and is proportional to the difference between its optimal velocity and its current real velocity $\dot x_k$.\par
\begin{figure}[hbt]
\begin{center}
  \includegraphics[scale=0.8]{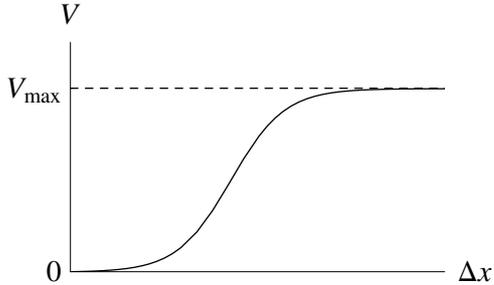}
\end{center}
\caption{Typical profile of optimal velocity.}
\label{fig:ov}
\end{figure}

  The typical profile of optimal velocity is shown in Fig.~\ref{fig:ov}.
This profile reflects a driver's behavior; if the distance from the vehicle ahead is short (long), he wants to keep low (high) speed.   When the distance becomes long enough, he wants to keep a speed limit of the road.  The results obtained by the optimal velocity model agree with real traffic data well.\par
  Nishinari and Takahashi reported an interesting relation between the Burgers equation and ECA184\cite{nishinari}.  They proposed a difference equation called `discrete Burgers equation' and showed that the Burgers equation and ECA184 are obtained by continuum and ultradiscrete limit respectively from the discrete Burgers equation.  Ultradiscretization is a method utilizing a non-analytic limit defined by the following formula\cite{tokihiro}.
\begin{equation}
  \lim_{\varepsilon\to+0}\varepsilon\log(e^{A/\varepsilon}+e^{B/\varepsilon}+\cdots)=\max(A,B,\cdots).
\end{equation}
We obtain a piecewise-linear type of equation called `ultradiscrete equation' by ultradiscretizing a difference equation.  There is a correspondence between basic operations of difference equation and of ultradiscrete one.  Usual operations $+$, $\times$ and $/$ of difference equation correspond to $\max$, $+$ and $-$ of ultradiscrete one respectively.  Thus we can make `analytic evaluation' for ultradiscrete equation as we do for difference equation.\par
  Moreover dependent variables can be discretized using appropriate initial data and constants of ultradiscrete equation.  Therefore ultradiscrete equation is a completely discretized equation in this sense.  Utilizing this feature, we can show that ECA184 originally defined by a binary table is equivalent to ultradiscrete Burgers equation.  Thus asymptotic behavior of solutions to ECA184 can be proved by the analytic evaluation reflecting that of Burgers equation.  As seen by this example, ultradiscretization gives a direct relation between CA and differential equation via difference one and proposes a new perspective for CA which can not be obtained if we make a closed analysis.\par
  In this letter, we propose a difference equation relevant to the OV model and call it `discrete OV (dOV) equation' .  If we take a continuum limit for this equation, we obtain (\ref{ov}) with a specific $V(\Delta x)$.  If we take an ultradiscrete limit, we obtain `ultradiscrete OV (uOV) equation' including ECA184 or FI model in a special case.  Since uOV equation is of second-order on time difference, it can express an acceleration effect.  Both dOV and uOV equations show a phase transition from free flow to jam.
\section{Discrete Optimal Velocity Model}
  Let us assume the same situation as of OV model (\ref{ov}).  The only difference is that a time variable is discrete.  Assume a time step denoted by $n$ ($n=0$, 1, $\cdots$) and an interval of time step by $\delta$ ($>0$).  Using these notations, dOV equation is defined by
\begin{equation}  \label{dov}
\begin{aligned}
 x_k^{n+1}-2x_k^n+x_k^{n-1}
= A\{&\log\bigl(1+\delta^2 V(x_{k+1}^n-x_k^n)\bigr) \\
&-\log\bigl(1+\delta(e^{x_k^n-x_k^{n-1}}-1)\bigr)\}.
\end{aligned}
\end{equation}
If $1+\delta^2 V(x_{k+1}^n-x_k^n)$ or $1+\delta(e^{x_k^n-x_k^{n-1}}-1)$  in the logarithmic terms is 0 or negative, (\ref{dov}) is not well-defined.  However, if $\delta$ is small enough and if $V(\Delta x)$ and initial data are appropriately defined, we can easily exclude this problem.\par
\begin{figure}[hbt]
\begin{center}
  \includegraphics[scale=0.8]{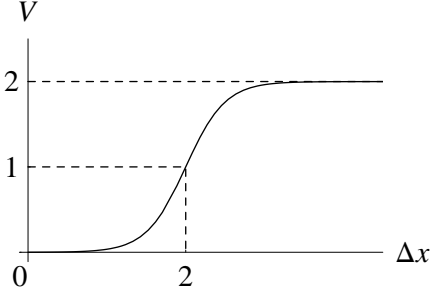}
\end{center}
\caption{Profile of $V(\Delta x)$ defined by (\ref{dov}) with $a=2$, $b=4$, $c=2$.}
\label{fig:dov}
\end{figure}
\begin{figure}[hbt]
\begin{center}
  \includegraphics[scale=0.8]{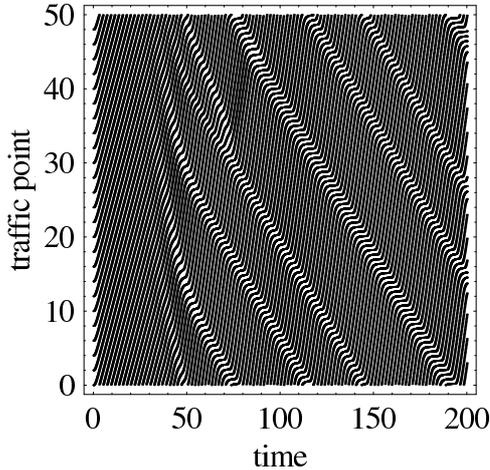}
\end{center}
\caption{Example of orbits of vehicles for $L=50$, $K=25$, $\delta=0.1$, $A=1$, $a=2$, $b=4$ and $c=2$.}
\label{fig:dov orbits}
\end{figure}

  Replacing $x_k^n$ by $x_k(n\delta)$ and assuming $\delta\sim0$, we obtain the following expansion.
\begin{equation}
  \ddot x_k=A\{V(x_{k+1}-x_k)-\dot x_k\}+\frac{A}{2}(\ddot x_k-(\dot x)^2)\delta+O(\delta^2).
\end{equation}
Thus (\ref{ov}) is derived from (\ref{dov}) by the continuum limit $\delta\to0$ and (\ref{dov}) is a discrete analogue to (\ref{ov}).  Considering this relation, $V(\Delta x)$ is required to have the profile roughly shown in Fig.~\ref{fig:ov}.  Moreover if we assume (\ref{dov}) can be ultradiscretized, $V(\Delta x)$ is required to have a more specific form.  To realize both continuum and ultradiscrete limit, we fix the following form for $V(\Delta x)$,
\begin{equation}  \label{dov V}
  V(\Delta x)=a\Bigl(\frac{1}{1+e^{-b(\Delta x-c)}}-\frac{1}{1+e^{bc}}\Bigr),
\end{equation}
where $a$, $b$ and $c$ are all positive constants.  Figure~\ref{fig:dov} shows an example of profile of $V(\Delta x)$.
Figure~\ref{fig:dov orbits} shows an example of orbits of vehicles.  Initial positions of vehicles are set at nearly regular intervals with small disturbances.  Coalescence of jams occurs at earlier time and three major jams survive in this figure.  Though not shown in this figure, more coalescences occur after a long time.\par
\begin{figure}[hbt]
\begin{center}
\setlength{\unitlength}{0.5cm}
\begin{picture}(10,9)(0,1)
  \put(0.6,5){(a)}
  \put(4.8,5.8){(b)}
  \put(7.7,3){(c)}
  \put(-3,10){\hbox to 0cm{\vbox to 0cm{\includegraphics[scale=0.9]{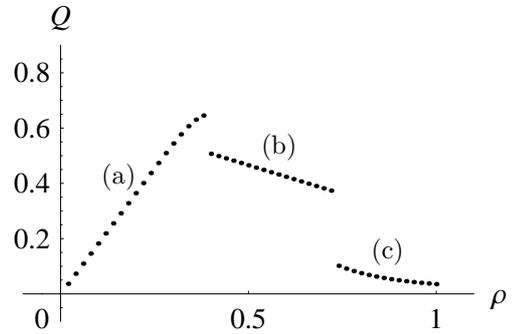}}}}
\end{picture}
\end{center}
\caption{Fundamental diagram with $L=50$, $\delta=0.1$, $A=1$, $a=2$, $b=4$, $c=2$, $n_0=90000$ and $n_1=100000$.  Plotted points are obtained for $1\le K\le 50$.}
\label{fig:dov fd}
\end{figure}

The fundamental diagram is shown in Fig.~\ref{fig:dov fd}\cite{chowdhury}.  This diagram shows a dependence of flow $Q$ on density $\rho$.  Density $\rho$ is a number of vehicles per unit length and flow $Q$ is equivalent to a total momentum of vehicles per unit length.  Both are defined by
\begin{equation}
\begin{aligned}
  \rho&=\frac{1}{L}(\text{number of vehicles}), \\
  Q&=\frac{1}{(n_1-n_0+1)L\delta}\sum_{n=n_0}^{n_1}\sum_{k=1}^K (x_k^{n}-x_k^{n-1}).
\end{aligned}
\end{equation}
We can observe three phases, that is, (a) free flow phase in a low density region, (b) jam phase in a medium density region and (c) tight jam phase in a high density region.  Since these phases can be observed for the OV model (\ref{ov}), we can consider that discretization of time variable in the dOV model (\ref{dov}) works well.
\section{Ultradiscrete Optimal Velocity Model}
  The dOV equation (\ref{dov}) with optimal velocity (\ref{dov V}) can be ultradiscretized.  Let us introduce transformation of variable and constants including a new parameter $\varepsilon$ defined by
\begin{equation}  \label{uov trans}
  x_k^n\to\frac{x_k^n+n\delta}{\varepsilon}, \quad
  \delta\to e^{-\delta/\varepsilon}, \quad
  a\to e^{(a+2\delta)/\varepsilon}, \quad
  c\to\frac{c}{\varepsilon}.
\end{equation}
Substituting the transformation into (\ref{dov}) and (\ref{dov V}), we obtain
\begin{equation}
\begin{aligned}
 & x_k^{n+1}-2x_k^n+x_k^{n-1} \\
={}& A\Bigl\{\varepsilon\log\Bigl(1+\frac{e^{a/\varepsilon}}{1+e^{-b(x_{k+1}^n-x_k^n-c)/\varepsilon}} - \frac{e^{a/\varepsilon}}{1+e^{bc/\varepsilon}}\Bigr) \\
&\qquad -\varepsilon\log\bigl(1+e^{(x_k^n-x_k^{n-1})/\varepsilon}-e^{-\delta/\varepsilon}\bigr)\Bigr\}.
\end{aligned}
\end{equation}
If $a$, $b$, $c$, $\delta$ are all positive and $a<bc$, we obtain the following ultradiscrete equation taking a limit $\varepsilon\to+0$.
\begin{equation}
\begin{aligned}
 & x_k^{n+1}-2x_k^n+x_k^{n-1} \\
={}& A\{\max(0,a-\max(0,-b(x_{k+1}^n-x_k^n-c)))\\
 & \qquad-\max(0,x_k^n-x_k^{n-1})\}.
\end{aligned}
\end{equation}
Moreover this equation is equivalent to
\begin{equation}  \label{uov}
\begin{aligned}
 & x_k^{n+1}-2x_k^n+x_k^{n-1} \\
={}& A\{V(x_{k+1}^n-x_k^n) -\max(0,x_k^n-x_k^{n-1})\},
\end{aligned}
\end{equation}
where
\begin{equation}  \label{uov V}
  V(\Delta x)=\max(0,b(\Delta x-c)+a)-\max(0,b(\Delta x-c)).
\end{equation}
Note that $\delta$ is not included in (\ref{uov}) due to the transformations of $x_k^n$ and $a$ in (\ref{uov trans}).\par
\begin{figure}[hbt]
\begin{center}
  \includegraphics[scale=0.8]{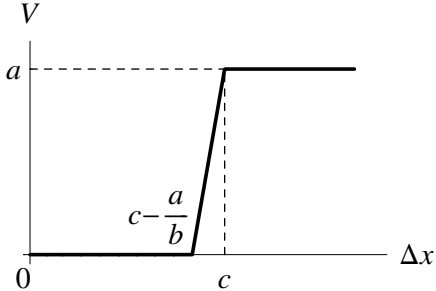}
\end{center}
\caption{Profile of $V(\Delta x)$ in (\ref{uov V}).}
\label{fig:uov}
\end{figure}
\begin{figure}[hbt]
\begin{center}
  \includegraphics[scale=0.8]{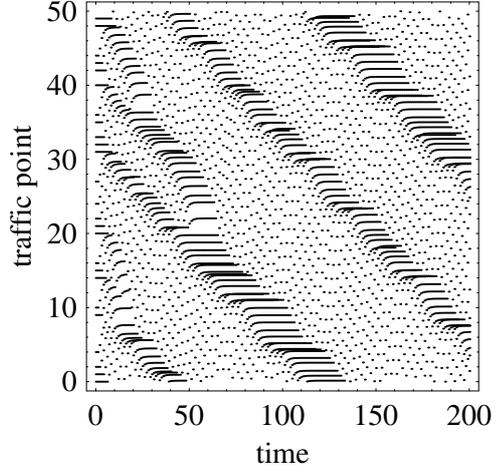}
\end{center}
\caption{Example of orbits of vehicles for (\ref{uov}) with $L=50$, $K=25$, $A=0.5$, $a=1.9$, $b=4$, $c=3$.}
\label{fig:uov orbits}
\end{figure}
\begin{figure}[hbt]
\begin{center}
  \includegraphics[scale=0.9]{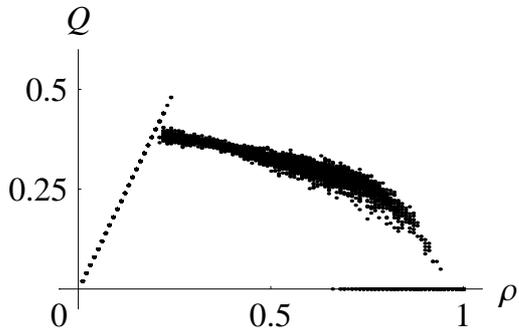}
\end{center}
\caption{Fundamental diagram for (\ref{uov}) with $L=100$, $A=0.5$, $a=1.9$, $b=4$, $c=3$, $n_0=1000$, $n_1=2000$.  Plotted points are obtained for $1\le K\le100$ and 50 trials are executed for every $K$.}
\label{fig:uov fd}
\end{figure}

  We show a typical profile of $V(\Delta x)$ in Fig.~\ref{fig:uov}.
Figure~\ref{fig:uov orbits} shows an example of orbits of vehicles.  Positions of vehicles are random integer at initial time step.  However $x_k^n$ is generally non-integer since $A$ and $a$ are not integer in this example.  A fundamental diagram using the same constants other than $K$ is shown in Fig.~\ref{fig:uov fd}.  Surprisingly three phases clearly exist as in Fig.~\ref{fig:dov fd}.  Note that numerical experiments are executed by double precision calculation of C program.
\section{Special Case of Ultradiscrete Optimal Velocity Model}
  In this section, we discuss two special cases of uOV model.
\subsection{No Overtaking}
  Assume $A$, $a$, $b$, $c$ in (\ref{uov}) and (\ref{uov V}) are all positive.  Then if $AV(\Delta x)\le \Delta x$, we can derive
\begin{equation}
  x_k^{n+1}\le x_{k+1}^n\underbrace{+x_k^n-x_k^{n-1}-A\max(0,x_k^n-x_k^{n-1})}_{(a)}.
\end{equation}
Moreover if $A\ge1$, (a) is always 0 or negative.  Therefore we get $x_k^{n+1}\le x_{k+1}^n$ on these assumptions.  And furthermore, if velocity of all vehicles is non-negative, overtaking does not occur.  When we use the OV (dOV, uOV) model as a numerical simulator of concrete traffic flow, overtaking of vehicle can not occur in a one-way circuit of single lane.  Though we can avoid overtaking by choosing appropriate constants and initial data, assurance of no overtaking is important for a real application.
\subsection{Cellular Automaton}
  If constants $A$, $a$, $b$, $c$ and initial position $x_k^0$ are all integer, any $x_k^n$ calculated by (\ref{uov}) is also integer.  Therefore dependent and independent variables in (\ref{uov}) are all discrete in this case.  Moreover, if we set $A=1$, (\ref{uov}) reduces to
\begin{equation}  \label{uov A=1}
  x_k^{n+1}=x_k^n+V(x_{k+1}^n-x_k^n)+\min(0,x_k^n-x_k^{n-1})
\end{equation}
Let us assume $V(\Delta x)\ge0$ as in Fig.~\ref{fig:uov}.  Moreover if $x_k^n-x_k^{n-1}\ge0$ for any $k$ at a certain $n$, the last term $\min(0,x_k^n-x_k^{n-1})$ in (\ref{uov A=1}) becomes 0 and $x_k^{n+1}-x_k^n=V(x_{k+1}^n-x_k^n)\ge0$ for any $k$.  Therefore any vehicle does not go backward if initial velocity of any vehicle is not negative.  Under this condition, (\ref{uov A=1}) again reduces to the first-order equation,
\begin{equation}  \label{uov A=1 v>=0}
  x_k^{n+1}=x_k^n+V(x_{k+1}^n-x_k^n).
\end{equation}
\par
  Moreover let us consider the case of $A=1$, $a=v_{\text{max}}$, $b=1$, $c=v_{\text{max}}+1$ where $v_{\text{max}}$ is positive integer.  Then $V(\Delta x)$ in (\ref{uov A=1 v>=0}) becomes
\begin{equation}
\begin{aligned}
 V(\Delta x)&=\max(0,\Delta x-1)-\max(0,\Delta x-v_{\text{max}}-1) \\
&=\begin{cases}
  0 & (\Delta x-1\le 0) \\
  \Delta x & (0<\Delta x-1<v_{\text{max}}) \\
  v_{\text{max}} & (v_{\text{max}}<\Delta x-1)
\end{cases}
\end{aligned}
\end{equation}
Assuming a size of vehicles is a unit cell size, $x_{k+1}^n-x_k^n-1$ is a distance between $k$-th and $(k+1)$-th vehicles at time step $n$.  Therefore every vehicle moves forward by its distance up to $v_{\text{max}}$.  This model is nothing but the FI model and ECA184 for $v_{\text{max}}=1$.  We note that an analogy between OV and FI models is commented by Nishinari\cite{nishinari2}.
\section{Concluding Remarks}
  We propose a new discrete OV model with a discrete time.  Continuum limit of this  model is equivalent to the OV model. We show that orbits of vehicles and the fundamental diagram agree with those of OV model qualitatively.  Moreover this model has an ultradiscrete limit and a piecewise-linear type of evolution equation is obtained.  We show that the ultradiscrete OV model also gives phase transition in its fundamental diagram by a numerical calculation.  It includes the FI model as a special case.\par
  We only show a definition, a few features and some numerical results about dOV and uOV models in this letter.  Detailed analysis using various combinations of constants is necessary to understand a dynamics of the models fully.  Comparison with other models and with real data is also necessary.  These points are future problems to be solved.
\renewcommand{\baselinestretch}{1.0}

\end{document}